# Journals Titles and Mission Statements – Lexical structure, diversity and readability

Julián David Cortés-Sánchez

**Affiliations:**

- Principal Professor – School of Management and Business, Universidad del Rosario, Colombia
- Invited Researcher – Fudan Development Institute, Fudan University, China

E-mail: julian.cortess@urosario.edu.co

Address: Calle 200 con autopista norte, Bogotá, Colombia

# Journals Titles and Mission Statements – Lexical structure, diversity and readability

**Abstract**

There is an established research agenda on dissecting an article's components— title and abstract readability and diversity, keywords, number references— and determining their association with bibliometrics performance. Yet, journals' titles and their overview, aim and scope (i.e., journal's mission statement – JMS(s) have not been investigated with the same diligence. This study aims to conduct a comprehensive outlook of titles and JMS's lexical structure and identify significant differences between journals' prestige and type of access groups and their JMS content in the field of business, management and accounting (BMA). Lexical network analysis was used to explore journals' title structure. JMS were examined through the Flesch-Kincaid grade level for readability and the Yule's K for lexical diversity. Titles and JMS structural analysis reflected current and critical discussion in BMA: an *obsession* for counterintuitive findings and ICT tools. JMS expressed mostly target customers and markets. JMS from reputable journals showed a higher betweenness for key-terms related to rigorous features, while JMS of lower reputable journals highlighted indexing attributes (i.e., Scopus). Wilcoxon rank-sum and Kruskal-Wallis tests showed significant differences in the JMS median diversity regarding the journal's type of access and best quartiles.

**Keywords:** Journals titles; Mission statement; Readability; Lexical diversity; Social Network Analysis

## 1 Introduction

How should we name our new journal? What could be the title, statement and scope differentiation content to brief our audience? Are those features described clearly and richly enough? Bottom line, does all of that have any association with bibliometric performance (output and citations)? Leading journals (e.g., *Scientometrics, Research Evaluation,* or *Academy of Management Journal*) possess a cemented *cumulative advantage* (Merton, 1968, 1988; Price, 1976), and these questions are not vertical priorities for their editors and administrators. Granted, space to publish in such journals is scarce (Bornmann & Marx, 2012) more so for researchers outside premier institutions in developing countries (Bravo et al., 2018; Dell'Anno et al., 2020; Smith et al., 2014). However, the landscape has been changing. Between 2019-2020 the bibliographic database Scopus added 820 new journals (Elsevier-Scopus, 2021; SCImago, 2020). Six percent of the journals were classified into the business, management, and accounting (BMA) subject area (Elsevier-Scopus, 2021; SCImago, 2020). Whilst new periodicals are emerging, readers and authors are eager to look around for journals with a clear and attractive title and statement/scope that eventually will catch or fit their attention, research interests and subsequent use (Rostami et al., 2014).

When it comes to choosing a journal likely to publish a researcher's work from developing countries, some shortcuts can be considered —looking at journal rankings (SCImago, 2020), indexation (Clarivate Analytics, 2021) or *manuscript-journal* matching services (e.g., Elsevier's or Wiley's journal finder; Springer's journal suggester). Masten and Aschcraft (2017) outlined the proper characteristics to avoid predatory and identify credible journals, such as i) intuitive website; ii) clear editor and editorial board information; iii) clear publisher information; iv) clear archiving and indexing; v) clear identification of the journal affiliation/sponsorship. This study focuses on the factors directly related to the questions outlined at the beginning of this introduction, namely an intuitive website, clear publisher information and journal characteristics, hereafter a journal's mission statement(s) (JMS) (Table 1).

[Table 1 about here]

In BMA, there is well-established research on mission statement(s) (MS) contents and its association with organisational performance (Bart, 1997, 2001; Bart & Baetz, 1998; Bart & Hupfer, 2004; Bart & Tabone, 2000; Cortés-Sánchez, 2018; Cortés-Sánchez & Rivera, 2019; Desmidt et al., 2011; Fitzgerald & Cunningham, 2016; Gharleghi et al., 2011; Godoy-Bejarano & Tellez-Falla, 2017; Palmer & Short, 2008; Pandey et al., 2017; Williams, 2008). The MS content should express at least four key concepts, namely purpose (why the organisation exists), values (what it believes in), standards and behaviours (the rules/norms that shape its operations), and strategy (long-term planning and pathway for achieving its purpose) (Campbell, 1989). Those key concepts are consistently aligned with those of the JMS. Table 1 proposes how MS and JMS are interrelated (e.g., The *Intuitive website* content characteristic of the JMS is related with the *Purpose* concept of the MS, since in it "the journal scope of content is well-defined and aligns with journal's title, article content and expectation of typical professional readers of the journal, instead of containing a broad scope of content").

Among the findings related to MS and organisational performance, it could be highlighted that MS with central content features, such as readability (i.e., clarity, brevity), lexical diversity, among others (e.g., mention of values/beliefs, social responsibility, lack of financial goals, positive language) are positively related to financial-innovation performance in organisations (Bart & Baetz, 1998; Bartkus et al., 2006; Berbegal-Mirabent et al., 2019; Cortés-Sánchez & Rivera, 2019; Duygulu et al., 2016; Godoy-Bejarano & Tellez-Falla, 2017; Zhang et al., 2015).

In the field of informetrics, on the other hand, there is an established agenda that seeks to disentangle the relationship between articles title, keywords, or abstract content (i.e., length, *informativeness*, accuracy, readability, presence of colons/question marks, changes over time) and their properties (i.e., length of the article, number of tables/figures), number of authors and performance (i.e., downloads or citations count) (Aleixandre-Benavent et al., 2014; Didegah & Thelwall, 2013; Haslam et al., 2008; Jamali & Nikzad, 2011; Lewison & Hartley, 2005; Li & Xu, 2019; Méndez et al., 2014; Rostami et al., 2014; Sahragard & Meihami, 2016; Uddin & Khan, 2016; Yitzhaki, 1994, 1997, 2002).

However, journals' titles and JMS are overlooked by that stream, despite their importance in describing a journal's main topics, history, disciplinary incidence, publication frequency, preferred epistemology (i.e., empirical or theoretical) and methods, alternative scholarly communications (e.g., letter to editors, essay, reviews), audience, geographical scope, membership, among other features. Furthermore, little research has been conducted related to readability and lexical diversity regardless of their importance for improving understanding performance and reflecting credibility and communication competence in written messages, vertical assets for emerging journals (Dillard & Pfau, 2012; Linsley & Lawrence, 2007; Rennekamp, 2012).

This study's aims are twofold. First, to conduct a comprehensive outlook of the content structure of titles and JMS. Second, to identify significant differences between journals' prestige and type of access groups and JMS content in readability and lexical diversity terms. It considers journals in BMA given the relevance and experience in the processes of researching, constructing and implementing MS. Semantic networks will be used for analysing the structure of titles and JMS. Readability and lexical diversity metrics will be used for analysing JMS attributes. Titles were excluded from the latter analysis considering that titles per se are too small for exploring their lexical diversity or readability compared to JMS. On that count, this study contributes to a structural assessment of journal titles and JMS quality in terms of readability and lexical diversity. It also adds to the literature on strategic planning and scholarly communication meta-knowledge (Evans & Foster, 2011). Results and open-access datasets could be useful for research institutions and editorials strategic planning activities related to the launch of new journals or improvement in the JMS of ongoing periodicals. Following this introduction, the article presents the methodology (data, methods and software). Then, it presents the discussion of the results. Finally, it presents the conclusion, restrictions and further research agenda.

## 2 Methodology

### 2.1 Data

Scopus was chosen over Web of Science (WoS) due to its broader journal coverage and the participation of researchers from developing countries (Baas et al., 2020; Mongeon & Paul-Hus, 2016). The SCImago journal ranking (2020) and the Scopus source list (2021) were chosen as a base-line list for collecting the JMS and additional information (Table 2). The journals considered for the final sample were those active as of 2019 indexed in at least one of the BMA categories (i.e., accounting; business and international management; industrial relations; management information systems; management of technology and innovation; marketing; organisational behaviour and human resource management; strategy and management; tourism, leisure and hospitality; and miscellaneous). A group of three research assistants curated by hand the JMS from the journals' website. Websites had either an overview (12.7%), additional information (e.g., journal's aim, mission, objective, scope) (47%), or both (40.3%). When the two options related to the JMS were available, both were sourced and unified in a single JMS. A total of 1,502 JMS were collected.

The SCImago Journal Rank (SJR) best quartiles were considered a proxy for journals' prestige and bibliometric performance (Guerrero-Bote & Moya-Anegón, 2012). Type of access was classified in two status: registered either in the Directory of Open Access Journals (DOAJ) or the Directory of Open Access Scholarly Resources (ROAD) or non-registered (DOAJ, n.d.; ROAD, n.d.), henceforth, OA and non-OA. Journals' type of access was considered as a crucial group considering the current structural changes in the scholarly communication dynamics due to the open access-science agenda (Chavarro et al., 2018; Jamali et al., 2016; Mayr, 2006). Table 2 presents the summary of variables considered. The complete dataset is available in the following permanent link: **[available upon request].**

[Table 2 about here]

### 2.2 Methods and software

For the titles and JMS structure, semantic network analysis was used. For the JMS content analysis (i.e., readability and lexical diversity) the Flesch-Kincaid grade level and Yule's K were employed. There are four types of networks: social, technological, biological and informational (Newman, 2003). The semantic networks here structured belong to the latter type. Semantic networks aim to examine the relationship among symbols' shared meaning (Doerfel & Barnett, 1999). First, each journal's title and JMS were passed to tokens (i.e., "[T]his or that word on a single line of a single page of a single copy of a book" (Peirce, 1906, pp. 505–506)). Punctuation, stop words and little-informative words (e.g., journal) were removed. A co-occurrence matrix was assembled to compute the number and direction of co-occurrences of tokens (e.g., in the last phrase, the token "co-occurrence(s)" co-occurred twice with the tokens "matrix" and "tokens"; both co-occurrences were directed from "co-occurrence(s)" to "matrix" and "tokens"). Finally, a directed-weighted semantic network was assembled based on the co-occurrence matrix. As a result, one can formally explore both the titles and JMS's key-terms underlying structure at a macro, meso and micro level with a diverse set of metrics.

At the macro level, the metrics considered were density and average path length, at the meso level, modularity, and, at the micro level, betweenness centrality (Wasserman & Faust, 2003). Density indicates how connected the titles and JMS key-terms are (i.e., the number of connections that exist compared to the number of potential connections that could exist), and the average path length illuminates the average number of steps along the shortest path for every pair of nodes (i.e., key-terms) in a network (Brandes, 2001; Iacobucci et al., 2018). It can also be understood as a measure of information flow efficiency (i.e., the closer to 1, the more efficient). Regarding the meso metrics, modularity indicates if there is a community-type structure in the network and the number of clusters formed (i.e., the higher the modularity, the denser the connections among nodes within clusters, and the sparse connection between nodes in other clusters) (Blondel et al., 2008; Traag et al., 2019). A higher betweenness score unveils a node's capacity to enable information flow between clusters (Opsahl et al., 2010).

Readability indices compute the clarity of texts intending to drop unnecessary complexities. One of the most used indices is the Flesch-Kincaid grade level (FKGL) (Sattari et al., 2011). Rudolf Flesch and J. Peter Kincaid developed it to assess the United States Navy's technical manuals' readability. In essence, it approximates the American school equivalent grade needed to comprehend a given text at the first reading (Kincaid et al., 1975). For instance, this paragraph FKGL is 9, which means that a person of about 15-16 years would understand this paragraph with no difficulty. The equation for the FKGL calculation is:

$$FKGL = 0.39 \left(\frac{words}{sentences}\right) + 11.8 \left(\frac{syllables}{words}\right) - 15.59$$

Source: Kincaid et al. (1975).

Based on Twedie and Baayen's (1998) review on the assessments of the lexical richness of texts, the Yule's K was considered for analysing the lexical diversity (Yule, 1944). The larger the Yule's K result, the more words have been repeated and less vocabulary richness was detected in a text. Conversely, the lower the results, the higher the lexical diversity. For instance, one of the highest JMS Yule's K results in the sample (*n*=1,428) (i.e., one of the lowest lexical diversity) is the following: "The International Journal of Educational Organization and Leadership inquiries into the nature and processes of effective educational administration and leadership." The equation for calculating Yule's K is:

$$K = 10^4 \times \left[-\frac{1}{N} + \sum_{i=1}^{V} f_v(i, N) \left(\frac{i}{N}\right)^2\right]$$

Source: Yule (1944).

Where $N$ refers to the total number of tokens, $V$ to the number of types (i.e., unique tokens), and $f_v(i, N)$ to the number of types occurring $i$ times in a sample of length $N$. The text corpora were processed using R's packages quanteda and igraph; Gephi was used for networks' layout and metrics computation (Bastian et al., 2009; Benoit et. al., 2020; R Core Team, 2014; The igraph core team, 2019).

## 3   Results and discussion

Figure 1 presents the titles' semantic network layout, top-20 frequent key-terms and network metrics. A random directed-network with the same number of nodes (*n*=608; wiring probability: 0.05) was generated as a comparing benchmark (i.e., density=0.025; average path length=2.56; clusters=11; modularity=0.152). The same was done for the following two JMS networks. Compared to the random network, the titles network showed a lower density, similar average path length and higher modularity.

The average path length shows that any key-term would have to contact ≈3 key-terms on average to reach another key-term. Modularity algorithm (Blondel et al., 2008) identified 21 clusters, although three clusters grouped ≈60% of the key-terms: pink (32%) related to marketing and strategic and supply chain management; green (22.2%) related to economics and finance; and blue (13.3%) related to sustainability sciences. The modularity score suggests a higher interconnection within clusters and loose connections between clusters. Journals' titles contain each field's distinctive jargon/language and share it moderately with other BMA fields. Key-term bridges between clusters are not strictly from BMA fields, such as the cases of *engineering* (e.g., Nuclear Engineering and Design; Ecological Engineering; Journal of Water Process Engineering), *technology* (e.g., Bioresource Technology; Journal of Quality Technology; Journal of Hospitality and Tourism Technology) or *development* (e.g., Entrepreneurship and Regional Development; Energy for Sustainable Development; Career Development Quarterly). They seem to establish comprehensive dialogues between (sub)fields.

[Figure 1 about here]

Due to processing restrictions and contrasting JMS of highly reputable and emerging journals, Figures 2 and 3 present JMS analysis of the top and bottom ten percent journals according to their SJR. A random directed-network with the average number of nodes of the following two networks ($n$=3,580; wiring probability: 0.05) was generated as a comparing benchmark (i.e., density=0.025; average path length=2.12; clusters=10; modularity=0.061). First, the top ten percent JMS showed similar density, average path length and number of clusters compared to the random network, although it showed higher modularity. The bottom ten percent JMS network metrics were similar to that of the top ten percent. In sum, both networks' macro scores were not different from those of a random network; however, they both differentiated themselves with a higher modularity score.

Both JMS networks' density and average path lengths were similar. In that line, any key-term of both networks would have to contact ≈2 key-terms on average to reach another key-term (e.g., there are, on average, two "middlemen" between two random type tokens). The number of clusters identified was 9 and 10, respectively. Regarding the top ten percent JMS network, the three most crowded grouped 67% of the key-terms: blue (31%) related to what the journals seek in manuscripts' (e.g., new/original/leading/innovative, conceptual/theoretical/empirical, rigorous); pink (28%) related to the professional/practical applications/methods (e.g., modeling, methodology, relevance, demonstrate) and audiences (e.g., students, professionals, policy-makers, leaders, non-governmental, scientists, researchers); and grey (10%) related to the social-environmental and geographical reach of the research (e.g., ecosystems, socioeconomic, municipal, communities). On the other hand, in the bottom ten percent JMS the three most crowded grouped 52% of the key-terms: pink (22%) related to what the journals seek in manuscripts' (e.g., original/case/wide/important/innovative); green (16%) related to journal attributes/focus and methodological preferences (e.g., peer-reviewed, quantitative/philosophical/multi-disciplinary); and blue (15%) related to private institutions processes/results (e.g., business, industry(ies), production/products/innovations, trade).

The modularity score of the top ten percent was lower than the bottom ten JMS. Therefore, there is a loose interconnection within clusters and higher interconnectedness between clusters. In other words, the JMS jargon/language in the top ten percent are broadly shared with other JMS (i.e., JMS showed no systematic differentiation when expressing what JMS seek, and research applications and audiences). Conversely, the modularity score of the bottom ten percent JMS suggests a higher interconnectedness within clusters and more loose between clusters. Thus, JMS jargon/language is also shared with other JMS (i.e., JMS showed a moderate differentiation when expressing what JMS seeks and journals' attributes and methodological preferences).

In the top ten percent JMS, key-term bridges between clusters were related to a local scope (e.g., *region, communities, national and international levels*); and critical type of scholarly communication (e.g., *critical evaluation, critical management studies, critical essays and other types of communication*s). Both JMS networks share the key-term: *life. Life* was a widely used key-term, related from biological sciences (e.g., life sciences) to marketing-related concepts (e.g., *product life cycle*). In the case of the bottom ten percent JMS, the key-terms with the highest betweenness were *year* (e.g., the advance of the discipline in a temporal framework or the journal's coverage or timely contribution to the field); *interested*, as in the interests of the journal regarding a manuscript attribute or the public interested (e.g., manuscripts that *break* the field; *state of the art* research; academics and practitioners as target audience); and *learn* (e.g., learn from the discipline or when the disciplines learn from other actors). It is crucial to notice that the indexing (i.e., *Scopus*) is a critical intermediate feature in the JMS of the bottom ten percent.

[Figure 2 about here]

Table 3 presents the summary statistics for JMS FKGL and Yule's K. Overall, the readability and lexical diversity of JMS is considerably lower. There was a small-significant correlation between FKGL and Yule's K (r=0.1; n=1,502; p=0.00006). Thus, increases in readability were positively —although slightly— correlated with increases in lexical diversity. The median FKGL indicates that 50 percent of the JMS have a readability score higher than 19, which means that a person would need ≈19 years equivalent to the US grade level to understand a text's first reading. The median Yule's K shows that 50 percent of the JMS have a higher lexical diversity score than ≈222. Compared to best-selling books such as the Harry Potter novels, which have a Yule's K of 69.38-77.12 (Reilly, 2020), JMS showed poor lexical diversity.

[Figure 3 about here]

Figures 4 and 5 present two world maps showing the median FKGL and Yule's K of the JMS for 54 countries according to the publishers' country with available data. As expected, there is consistency in the readability of JMS in journals that the publisher is from the English-speaking world (e.g., England, Ireland, USA, Canada, Australia, New Zealand), including other countries such as Sweden, Italy, Pakistan, Iran or Japan. On the other hand, Iberoamerican countries (e.g., Spain, Portugal, Colombia, Brazil, Mexico) are among the countries with the lowest readability. Fewer JMS from non-English speaking countries showed a higher lexical diversity (e.g., Colombia, India, Finland) than those from the English-speaking world (e.g., Australia). In both categories, the USA and Canada are among the few countries with higher readability and lexical diversity. Conversely, there are JMS from countries with both lower readability and lexical diversity (e.g., China, Russia, Ukraine, Spain).

[Table 3 about here]

[Figure 4 about here]

[Figure 5 about here]

None of the Shapiro-Wilk tests for FKGL and Yule's K conducted for each OA and best quartile groups resulted in a $p>0.05$ (Shapiro & Wilk, 1965). Considering that FKGL and Yule's K were not normally distributed, the Wilcoxon rank-sum and the Kruskal-Wallis tests were used to determine significant differences between OA and non-OA, and best quartiles (i.e., Q1, Q2, Q3, Q4) groups in terms of readability and lexical diversity (Hettmansperger & McKean, 1998). Results showed that the median FKGL in the OA group was 19.6 (IQR=6.53), whereas the median in the non-OA group was 19.5 (IQR=5.16). The Wilcoxon test showed that the difference was non-significant (p=0.19; effect size r=0.03). On the other hand, the median Yule's K in the OA group was 229 (IQR=110), whereas the median in the non-OA group was 222 (IQR=89.8). The Wilcoxon test showed that the difference was significant (p=0.02; effect size r=0.05).

There were statistically significant differences between the best quartile groups in terms of JMS readability (p=0.006). Pairwise Dunn's test between best quartile groups confirmed that the difference between Q2 (median=20; IQR=5.4) and Q4 (median=19; IQR=6.13) groups was significant (p=0.0005, eta2[H]=0.006). There were also statistically significant differences between best quartile groups in JMS Yule's K (p=0.000000325). Pairwise Dunn's test between best quartile groups showed significant differences between Q1 (median=214; IQR=84) and Q3 (median=227; IQR=92.2) (p=0.017, eta2[H]=0.006); and Q1 and Q4 (median=246; IQR=125) (p=0.000000148; eta2[H]=0.006). Also, between Q4 and Q2 (median=222; IQR=87.8) (p=0.000575; eta2[H]=0.006); and Q4 and Q3 (median=227; IQR=92.2) (p=0.0195; eta2[H]=0.006) groups.

The titles network structure showed a lower density, which unveils an open-type network structure. That was also noticeable in the JMS networks. An open-type structure has its advantages and disadvantages. Open structure networks enable the flow of diverse, new and non-redundant information (Yang et al., 2020). Closed structure networks are information-redundant but more efficient when extending control and implementing sanctions (Stokman, 2001). That reflects current discussions on the purpose of BMA-related research, regarding an incentive system that rewards novelty (i.e., non-redundancy) over *truth* (i.e., heading towards an accumulated and progressive consensus) and apply few punishments for counterintuitive but erroneous findings (Davis, 2015). The higher betweenness centrality of key-terms such as *engineering* and *technology* may be associated with the increasing role of information and communication technologies (ICT) tools in organisations processes and research methods (e.g., Big Data), particularly in the cluster related to marketing and strategic and supply chain management (Davis, 2010, 2015; George et al., 2014). The fraction of key-terms composing the sustainability science-related cluster support the reported success of sustainability science concepts since the 80s, its explosive growth, the diverse geographical origin of its contributors and its emphasis on management (e.g., of social or ecological ecosystems) (Bettencourt & Kaur, 2011).

JMS subsamples showed that the key components (i.e., most crowded clusters) for a JMS were: i) what the journals seek —and want to publish, in a way— in submitted manuscripts; ii) professional/practical applications/methods; iii) audiences; iv) the social-environmental and geographical reach of the research; and v) journal's topics associated with mostly private institutions processes/results. Seminal work by Pearce and David (1987) proposed eight key components of a comprehensive MS: i) the specification of target customers and markets; ii) the identification of principal products/services; iii) the specification of a geographic domain, iv) the identification of core technologies; v) the expression of commitment to survival, growth and profitability; vi) the specification of key elements in the company's philosophy; vii) the identification of the company self-concept and viii) the identification of the firm's desired public image.

Contrasting those key components with those found in the JMS networks, JMS are mainly reflecting four Pearce and David (1987) key components: i) the specification of target customers and markets is associated with the a) audiences and b) journal's topics associated with mostly private institutions processes/results components; ii) the identification of principal products/services to offer (i.e., articles content) is associated with the component related to what the journals seek in submitted manuscripts; iii) the identification of a geographic domain is associated with the social-environmental and geographical reach of the research; and iv) the identification of core technologies (i.e., means for ends) is associated with the professional/practical applications/methods. There are several key MS components sidelined from the main JMS components detected. To synthesise, JMS could explore key components such as how the journal plans to grow and maintains its relevance and sustainability; what is its main reason/purpose as philosophy, self-concept and the public image than it is developing or seeks to achieve in the future (Masten & Ashcraft, 2017, p. 5).

Key-terms with higher betweenness associated with identifying principal products/services to offer such as critical type and other attributes of scholarly communication (i.e., *critical, interested*) also reflect a current fixation of the BMA related fields with *interesting theories* and *counterintuitive* facts (Davis, 1971). Pillutla and Thau (2013) argued that that focus had produced adverse effects: since the priority is to show something interesting, there is no incentive or mandate to share data for replication or triangulation, let alone expending resources testing intuitive phenomena to head towards a consensus or theory robustness (Aarts et al., 2015; Baker & Penny, 2016; Chawla, 2016). Apart from that, there is a noticeable discrepancy among the most frequent and highly intermediate key-terms. *International* can be found among the top-5 most frequent terms in the titles and JMS networks, and, at the same time, *local* was the key-term with the highest betweenness among the top ten percent JMS. *Local's* betweenness reflects an inherent feature of the social sciences: it is local in scope and relies on national publishers (Larivière et al., 2015).

Furthermore, an exploratory search on Google's Books Ngram Viewer shows that *international* and *local* —and other related terms— have increased jointly between 1800-1980 (Figure 6). Although, since the mid-1990s, their use dropped. Other terms (e.g., *global*), meanwhile, started to emerge (Google, 2021). The betweenness score and frequent use of a *local* and *international* agenda seem to be replaced by a global perspective when taking a glimpse of a broader context (e.g., *global and: cross-cultural issues; competitiveness; environmental change; political economy; supply chain*).

[Figure 6 about here]

The top ten percent JMS highlighted the betweenness of *rigorous* (e.g, *research; articles; contributions; peer review; sound theory*), which was not detected among the bottom ten percent JMS. Both JMS subsamples mentioned the journal's indexing, yet mentioning *Scopus* and *years* in the bottom ten percent JMS was a central attribute. Leading journals do not need to announce their indexing since their historical reputation is widely acknowledged. Emergent journals seem to use these indexing and timely factors as *hooks* for contributors and could be associated with the MS key component to identifying the firm's desired public image (Pearce II & David, 1987). Institutions also incentivise researchers to publish in Scopus' indexed journals — excessively, in some cases— creating structural problems in the quality of BMA research output in developing countries (Cortés-Sánchez, 2019, 2020). The higher betweenness of *learn* in the bottom ten percent JMS supports its high frequency in the MS of universities worldwide, particularly in the small ones (Cortés-Sánchez, 2018).

Despite the relatively higher readability and diversity of JMS from countries from the English-speaking world, results are consistent with those arguing that academic and business and economics-related written pieces are far from a text understandable by a general audience (Cortés-Sánchez & Barredo, 2020; Stone & Lodhia, 2019). After analysing 709,000+ research papers abstracts, Plavén-Sigray et al. (2017) found that science's readability decreases overtime. As for 2015, 22% of abstracts have readability beyond college graduate English-level.

There are no significant differences in the readability of the JMS of OA and non-OA. That contrast findings in OA journals in psychology, where Plan Language Summaries (PLS) (i.e., non-technical language briefs for a wider audience) are being used accompanying the traditional abstracts, which happens to have a significantly higher readability than traditional abstracts (Stricker et al., 2020). Latter results shall be interpreted cautiously since only two journals that used PLS were assessed. By comparison, the median lexical diversity of JMS in the non-OA group was significantly different and higher than the OA group. Most of the publishers that compose the non-OA are those associated with the academic publishing oligopoly, such as Elsevier or Taylor & Francis (Larivière et al., 2015). Therefore, the JMS of for-profit publishers reflects the experience in using lexical diversity in strategic communications.

JMS median readability was significantly different and higher only for Q4 journals than Q2. It supports Didegah and Thelwall's (2013) findings on no significant association between abstracts readability and impact. A reason argued is that academics usually write abstracts for a specialised audience, even more in medicine and chemistry, while mathematics and space sciences have higher readability (Gazni, 2011). The median JMS lexical diversity was significantly different and higher in Q1 and Q2 journals than in Q3 and Q4 journals. Those findings support previous research from multiple disciplines that examined the increasing lexical diversity of article titles to address higher communicative attributes and catching readers' attention in leading journals, likewise, the diversification of titles from articles keywords and its association with higher impact (Méndez et al., 2014; Rostami et al., 2014; Yitzhaki, 2002). Previous results where titles attribute such as *catchiness* were not associated with impact should be contemplated concerning research restrictions (i.e., titles were assessed manually and subjectively) (Haslam et al., 2008).

## 4 Conclusion

Journals' titles and JMS are two essential information pieces that express a journal's main topics, history, disciplinary incidence, publication frequency, preferred epistemology (i.e., empirical or theoretical), methods, audiences, among other components. This study conducted a comprehensive outlook of the content structure of titles and JMS and identified significant differences between journal prestige and journal type of access groups and JMS content. Findings showed that both titles and JMS have an open-type network structure, enabling the flow of new information yet making that process less efficient than a closed structured network. Both titles and JMS reflect current discussions inside BMA (e.g., a *that is interesting* and *counterintuitive* findings obsession; increasing use of ICT tools in methods and industry; and a widespread incursion of sustainability sciences). Compared with the MS literature, JMS were restricted to the specification of target customers and markets; identifying principal products/services to offer; identifying a geographic domain; and identifying core technologies, leaving aside several MS components, such as how the journal plans to grow and maintain its relevance and sustainability, its main reason/purpose as philosophy, self-concept and the public image that it is developing or seeks to achieve in the future.

Reputable journals showed a distinctive key-term betweenness related to articles' *rigorous* features, while the publication hook in lower reputable JMS was to be indexed in *Scopus*. There is a tension between the frequent use of *international* and *local* betweenness. In the broader framework, both *international* and *local* are being less used over the past 20 years, while the notion of *global* is increasing. Both titles and JMS readability and lexical diversity are low, a phenomenon confirmed in multiple and interdisciplinary studies. However, the group of leading editorials composing the non-OA group showed JMS with higher lexical diversity. There were no significant differences between best quartile groups in terms of readability, but several significant differences emerge when exploring lexical diversity. Results are supported by previous research conducted for understanding the diversity of article titles and their association with impact. This study's restrictions could rely on readability and diversity indices' limitations since they cannot differentiate each reader. Further studies could source journals' information from different sources (e.g., WoS; Dimensions; Google Scholar) and disciplines for comparison (e.g., BMA vs. STEM disciplines). Also, adding additional article components, such as titles, abstracts, keywords, number of figures/tables, references, length, their readability and diversity and other text properties (e.g., sentiment dimensions: negative, positive, uncertain, litigious, constraining, superfluous, or interesting tones) and their potential associations with journal's use and impact in multiple spheres (e.g., views, downloads, altmetrics).

**References**

Aarts, A. A., Anderson, J. E., Anderson, C. J., Attridge, P. R., Attwood, A., Axt, J., Babel, M., Bahník, Š., Baranski, E., Barnett-Cowan, M., Bartmess, E., Beer, J., Bell, R., Bentley, H., Beyan, L., Binion, G., Borsboom, D., Bosch, A., Bosco, F. A., … Zuni, K. (2015). Estimating the reproducibility of psychological science. *Science*, *349*(6251). https://doi.org/10.1126/science.aac4716

Aleixandre-Benavent, R., Montalt-Resurecció, V., & Valderrama-Zurián, J. C. (2014). A descriptive study of inaccuracy in article titles on bibliometrics published in biomedical journals. *Scientometrics*, *101*(1), 781–791. https://doi.org/10.1007/s11192-014-1296-5

Baas, J., Schotten, M., Plume, A., Côté, G., & Karimi, R. (2020). Scopus as a curated, high-quality bibliometric data source for academic research in quantitative science studies. *Quantitative Science Studies*, *1*(1), 377–386. https://doi.org/10.1162/qss_a_00019

Baker, M., & Penny, D. (2016). Is there a reproducibility crisis? In *Nature* (Vol. 533, Issue 7604, pp. 452–454). Nature Publishing Group. https://doi.org/10.1038/533452A

Bart, C. K. K. (1997). Industrial firms and the power of mission. *Industrial Marketing Management*, *26*(4), 371–383. https://doi.org/10.1016/S0019-8501(96)00146-0

Bart, C. K. K. (2001). Exploring the application of mission statements on the World Wide Web. *Internet Research*, *11*(4), 360–368. https://doi.org/10.1108/10662240110402812

Bart, C. K. K., & Baetz, M. C. C. (1998). The relationship between mission statements and firm performance: An exploratory study. *Journal of Management Studies*, *35*(6), 823–853. https://doi.org/10.1111/1467-6486.00121

Bart, C. K. K., & Hupfer, M. (2004). Mission statements in Canadian hospitals. *Journal of Health Organization and Management*, *18*(2), 92–110. https://doi.org/10.1108/14777260410538889

Bart, C. K. K., & Tabone, J. C. C. (2000). Mission statements in Canadian not-for-profit hospitals: Does process matter? *Health Care Management Review*, *25*(2), 45–63. https://doi.org/10.1097/00004010-200004000-00005

Bartkus, B., Glassman, M., & McAfee, R. B. (2006). Mission statement quality and financial performance. *European Management Journal, 24*(1), 86–94. https://doi.org/10.1016/j.emj.2005.12.010

Bastian, M., Heymann, S., & Jacomy, M. (2009). Gephi: an open source software for exploring and manipulating networks. *International AAAI Conference on Weblogs and Social Media*. https://gephi.org/users/publications/

Benoit et. al. (2020). *Quantitative Analysis of Textual Data* (2.1.2). CRAN.

Berbegal-Mirabent, J., Mas-Machuca, M., & Guix, P. (2019). Impact of mission statement components on social enterprises' performance. *Review of Managerial Science*. https://doi.org/10.1007/s11846-019-00355-2

Bettencourt, L. M. A., & Kaur, J. (2011). Evolution and structure of sustainability science. *Proceedings of the National Academy of Sciences of the United States of America*, *108*(49), 19540–19545. https://doi.org/10.1073/pnas.1102712108

Blondel, V. D., Guillaume, J.-L., Lambiotte, R., & Lefebvre, E. (2008). Fast unfolding of communities in large networks. *Journal of Statistical Mechanics: Theory and Experiment*, *2008*(10), P10008.

Bornmann, L., & Marx, W. (2012). The Anna Karenina principle: A way of thinking about success in science. *Journal of the American Society for Information Science and Technology*, *63*(10), 2037–2051. https://doi.org/10.1002/asi.22661

Brandes, U. (2001). A faster algorithm for betweenness centrality. *Journal of Mathematical Sociology*, *25*(2), 163–177. https://doi.org/10.1080/0022250X.2001.9990249

Bravo, G., Farjam, M., Grimaldo Moreno, F., Birukou, A., & Squazzoni, F. (2018). Hidden connections: Network effects on editorial decisions in four computer science journals. *Journal of Informetrics*, *12*(1), 101–112. https://doi.org/10.1016/j.joi.2017.12.002

Campbell, A. (1989). Does your organisation need a mission? *Leadership & Organization Development Journal*, *10*(3), 3–9. https://doi.org/10.1108/EUM0000000001134

Chavarro, D., Ràfols, I., & Tang, P. (2018). To what extent is inclusion in the Web of Science an indicator of journal "quality"? *Research Evaluation*, *27*(2), 106–118. https://doi.org/10.1093/reseval/rvy001

Chawla, D. S. (2016). How many replication studies are enough? *Nature*, *531*(7592), 11. https://doi.org/10.1038/531011f

Clarivate Analytics. (2021). *Web of Science Master Journal List*. https://mjl.clarivate.com/home

Cortés-Sánchez, J. D. (2018). Mission statements of universities worldwide - Text mining and visualization. *Intangible Capital*, *14*(4), 584–603.

https://doi.org/10.3926/ic.1258

Cortés-Sánchez, J. D. (2019). Innovation in Latin America through the lens of bibliometrics: crammed and fading away. *Scientometrics*, *121*(2), 869–895. https://doi.org/10.1007/s11192-019-03201-0

Cortés-Sánchez, J. D. (2020). A bibliometric outlook of the most cited documents in business, management and accounting in Ibero-America. *European Research on Management and Business Economics*, *26*(1), 1–8. https://doi.org/10.1016/j.iedeen.2019.12.003

Cortés-Sánchez, J. D., & Barredo, D. (2020). Content Analysis in Business Digital Media Columns: Evidence From Colombia. *Journalism Practice*. https://doi.org/10.1080/17512786.2020.1796762

Cortés-Sánchez, J. D., & Rivera, L. (2019). Mission statements and financial performance in Latin-American firms. *Business: Theory and Practice*, *20*, 270–283. https://doi.org/10.3846/btp.2019.26

Davis. (2010). Do Theories of Organizations Progress? *Organizational Research Methods*, *13*(4), 690–709. https://doi.org/10.1177/1094428110376995

Davis. (2015). What Is Organizational Research For? *Administrative Science Quarterly*, *60*(2), 179–188. https://doi.org/10.1177/0001839215585725

Davis, M. (1971). That's Interesting!: Towards a Phenomenology of Sociology and a Sociology of Phenomenology. *Philosophy of the Social Sciences*, *1*(2), 309–344. https://doi.org/10.1177/004839317100100211

Dell'Anno, R., Caferra, R., & Morone, A. (2020). A "Trojan Horse" in the peer-review process of fee-charging economic journals. *Journal of Informetrics*, *14*(3), 101052. https://doi.org/10.1016/j.joi.2020.101052

Desmidt, S., Prinzie, A., & Decramer, A. (2011). Looking for the value of mission statements: A meta-analysis of 20 years of research. *Management Decision*, *49*(3), 468–483. https://doi.org/10.1108/00251741111120806

Didegah, F., & Thelwall, M. (2013). Which factors help authors produce the highest impact research? Collaboration, journal and document properties. *Journal of Informetrics*, *7*(4), 861–873. https://doi.org/10.1016/j.joi.2013.08.006

Dillard, J. P., & Pfau, M. (2012). The Persuasion Handbook: Developments in Theory and Practice. In *The Persuasion Handbook: Developments in Theory and Practice*. SAGE Publications, Inc. https://doi.org/10.4135/9781412976046

DOAJ. (n.d.). *Directory of Open Access Journals (DOAJ)*. Retrieved February 11, 2021, from https://doaj.org/

Doerfel, M. L., & Barnett, G. A. (1999). A Semantic Network Analysis of the International Communication Association. *Human Communication Research*, *25*(4), 589–603. https://doi.org/10.1111/j.1468-2958.1999.tb00463.x

Duygulu, E., Ozeren, E., Işıldar, P., & Appolloni, A. (2016). The Sustainable strategy for small and medium sized enterprises: The relationship between mission statements and performance. *Sustainability (Switzerland)*, *8*(7). https://doi.org/10.3390/su8070698

Elsevier-Scopus. (2021). *Scopus Roadmap: What's coming up in 2020 & 2021?* https://blog.scopus.com/posts/scopus-roadmap-whats-coming-up-in-2020-2021

Evans, J. A., & Foster, J. G. (2011). Metaknowledge. *Science*, *331*(6018), 721

LP – 725. https://doi.org/10.1126/science.1201765

Fitzgerald, C., & Cunningham, J. A. (2016). Inside the university technology transfer office: mission statement analysis. *Journal of Technology Transfer*, *41*(5), 1235–1246. https://doi.org/10.1007/s10961-015-9419-6

Fruchterman, T. M. J., & Reingold, E. M. (1991). Graph drawing by force-directed placement. *Software: Practice and Experience*, *21*(11), 1129–1164. https://doi.org/10.1002/spe.4380211102

Gazni, A. (2011). Are the abstracts of high impact articles more readable? Investigating the evidence from top research institutions in the world. *Journal of Information Science*, *37*(3), 273–281. https://doi.org/10.1177/0165551511401658

George, G., Haas, M. R., & Pentland, A. (2014). From the editors: Big data and management. In *Academy of Management Journal* (Vol. 57, Issue 2, pp. 321–326). Academy of Management. https://doi.org/10.5465/amj.2014.4002

Gharleghi, E., Nikbakht, F., & Bahar, G. (2011). A survey of relationship between the characteristics of mission statement and organizational performance. *Research Journal of Business Management*, *5*(3), 117–124. https://doi.org/10.3923/rjbm.2011.117.124

Godoy-Bejarano, J. M. M., & Tellez-Falla, D. F. F. (2017). Mission Power and Firm Financial Performance. *Latin American Business Review*, *18*(3–4), 211–226. https://doi.org/10.1080/10978526.2017.1400389

Google. (2021). *Google Ngram Viewer*. https://books.google.com/ngrams

Guerrero-Bote, V. P., & Moya-Anegón, F. (2012). A further step forward in measuring journals' scientific prestige: The SJR2 indicator. *Journal of Informetrics*, *6*, 674–688. https://doi.org/10.1016/j.joi.2012.07.001

Haslam, N., Ban, L., Kaufmann, L., Loughnan, S., Peters, K., Whelan, J., & Wilson, S. (2008). What makes an article influential? Predicting impact in social and personality psychology. *Scientometrics*, *76*(1), 169–185. https://doi.org/10.1007/s11192-007-1892-8

Hettmansperger, T. P., & McKean, J. W. (1998). *Robust nonparametric statistical methods*. John Wiley and Sons Inc.

Iacobucci, D., McBride, R., Popovich, D. L., & Rouziou, M. (2018). In Social Network Analysis, Which Centrality Index Should I Use?: Theoretical Differences and Empirical Similarities among Top Centralities. *Journal of Methods and Measurement in the Social Sciences*, *8*(2), 72–99. https://doi.org/10.2458/v8i2.22991

Jamali, H. R., Nicholas, D., & Herman, E. (2016). Scholarly reputation in the digital age and the role of emerging platforms and mechanisms. *Research Evaluation*, *25*(1), 37–49. https://doi.org/10.1093/reseval/rvv032

Jamali, H. R., & Nikzad, M. (2011). Article title type and its relation with the number of downloads and citations. *Scientometrics*, *88*(2), 653–661. https://doi.org/10.1007/s11192-011-0412-z

Kincaid, J., Fishburne, R., Rogers, R., & Chissom, B. (1975). *Derivation of new readability formulas (automated readability index, fog count, and flesch reading ease formula) for Navy enlisted personnel*. Chief of Naval Technical Training.

Larivière, V., Haustein, S., & Mongeon, P. (2015). The Oligopoly of Academic Publishers in the Digital Era. *PLOS ONE*, *10*(6), e0127502. https://doi.org/10.1371/journal.pone.0127502

Lewison, G., & Hartley, J. (2005). What's in a title? Numbers of words and the presence of colons. *Scientometrics*, *63*(2), 341–356. https://doi.org/10.1007/s11192-005-0216-0

Li, Z., & Xu, J. (2019). The evolution of research article titles: the case of Journal of Pragmatics 1978–2018. *Scientometrics*, *121*(3), 1619–1634. https://doi.org/10.1007/s11192-019-03244-3

Linsley, P. M., & Lawrence, M. J. (2007). Risk reporting by the largest UK companies: Readability and lack of obfuscation. *Accounting, Auditing and Accountability Journal*, *20*(4), 620–627. https://doi.org/10.1108/09513570710762601

Masten, Y., & Ashcraft, A. (2017). Due diligence in the open-access explosion era: choosing a reputable journal for publication. *FEMS Microbiology Letters*, *364*(21). https://doi.org/10.1093/femsle/fnx206

Mayr, P. (2006). Constructing experimental indicators for open access documents. *Research Evaluation*, *15*(2), 127–132. https://doi.org/10.3152/147154406781775940

Méndez, D. I., Ángeles Alcaraz, M., & Salager-Meyer, F. (2014). Titles in English-medium Astrophysics research articles. *Scientometrics*, *98*(3), 2331–2351. https://doi.org/10.1007/s11192-013-1174-6

Merton, R. (1968). The matthew effect in science. *Science*, *159*(3810), 56–62. https://doi.org/10.1126/science.159.3810.56

Merton, R. (1988). The Matthew Effect in Science, II: Cumulative Advantage and the Symbolism of Intellectual Property. *Isis*, *79*(4), 606–623. http://www.jstor.org/stable/234750

Mongeon, P., & Paul-Hus, A. (2016). The journal coverage of Web of Science and Scopus: a comparative analysis. *Scientometrics*, *106*(1), 213–228. https://doi.org/10.1007/s11192-015-1765-5

Newman, M. E. J. (2003). The Structure and Function of Complex Networks. *SIAM Rev.*, *45*(2), 167–256.

Opsahl, T., Agneessens, F., & Skvoretz, J. (2010). Node centrality in weighted networks: Generalizing degree and shortest paths. *Social Networks*, *32*(3), 245–251. https://doi.org/10.1016/j.socnet.2010.03.006

Palmer, T. B. B., & Short, J. C. C. (2008). Mission statements in U.S. colleges of business: An empirical examination of their content with linkages to configurations and performance. *Academy of Management Learning and Education*, *7*(4), 454–470. https://doi.org/10.5465/AMLE.2008.35882187

Pandey, S. K., Kim, M., & Pandey, S. K. (2017). Do Mission Statements Matter for Nonprofit Performance?: Insights from a Study of US Performing Arts Organizations. *Nonprofit Management and Leadership*, *27*(3), 389–410. https://doi.org/10.1002/nml.21257

Pearce II, J. A., & David, F. (1987). Corporate Mission Statements: The Bottom Line. *Academy of Management Executive (08963789)*, *1*(2), 109–115. https://libproxy.berkeley.edu/login?qurl=http%3A%2F%2Fsearch.ebscohost.com%2Flogin.aspx%3Fdirect%3Dtrue%26db%3Dbth%26AN%3D4275821%26site%3Deds-live

Peirce, C. S. (1906). Prolegomena to an apology for pragmaticism. *The Monist*, *16*(4), 492–462.

Pillutla, M. M., & Thau, S. (2013). Organizational sciences' obsession with "that's interesting!" *Organizational Psychology Review*, *3*(2), 187–194. https://doi.org/10.1177/2041386613479963

Plavén-Sigray, P., Matheson, G. J., Schiffler, B. C., & Thompson, W. H. (2017). The readability of scientific texts is decreasing over time. *ELife*, *6*. https://doi.org/10.7554/eLife.27725

Price, D. D. S. (1976). A general theory of bibliometric and other cumulative advantage processes. *Journal of the American Society for Information Science*, *27*(5), 292–306. https://doi.org/https://doi.org/10.1002/asi.4630270505

R Core Team. (2014). R: A language and environment for statistical computing. In *R Foundation for Statistical Computing, Vienna, Austria.* (Vol. 0, pp. 1–2667). R Foundation for Statistical Computing. online: http://www. R-project. org

Reilly, J. (2020). *Harry Potter and the Lexical Diversity Measures*. https://reilly-lab.github.io/Jamie_LexDiversity.html

Rennekamp, K. (2012). Processing Fluency and Investors' Reactions to Disclosure Readability. *Journal of Accounting Research*, *50*(5), 1319–1354. https://doi.org/10.1111/j.1475-679X.2012.00460.x

ROAD. (n.d.). *Welcome to ROAD*. Retrieved February 11, 2021, from https://road.issn.org/

Rostami, F., Mohammadpoorasl, A., & Hajizadeh, M. (2014). The effect of characteristics of title on citation rates of articles. *Scientometrics*, *98*(3), 2007–2010. https://doi.org/10.1007/s11192-013-1118-1

Sahragard, R., & Meihami, H. (2016). A diachronic study on the information provided by the research titles of applied linguistics journals. *Scientometrics*, *108*(3), 1315–1331. https://doi.org/10.1007/s11192-016-2049-4

Sattari, S., Pitt, L. F., & Caruana, A. (2011). How readable are mission statements? An exploratory study. *Corporate Communications: An International Journal*, *16*(4), 282–292. https://doi.org/10.1108/13563281111186931

SCImago. (2020). *SJR : Scientific Journal Rankings*. https://www.scimagojr.com/journalrank.php

Scopus. (2021). *Scopus preview*. https://www.scopus.com/

Shapiro, S. S., & Wilk, M. B. (1965). An analysis of variance test for normality (complete samples). *Biometrika*, *52*(3–4), 591–611. https://doi.org/10.1093/biomet/52.3-4.591

Six, J. M., & Tollis, I. G. (2006). A framework and algorithms for circular drawings of graphs. *Journal of Discrete Algorithms*, *4*(1), 25–50. https://doi.org/10.1016/j.jda.2005.01.009

Smith, M. J., Weinberger, C., Bruna, E. M., & Allesina, S. (2014). The Scientific Impact of Nations: Journal Placement and Citation Performance. *PLoS ONE*, *9*(10), e109195. https://doi.org/10.1371/journal.pone.0109195

South, A. (2011). rworldmap: A New R package for Mapping Global Data. *The R Journal*, *3*(1), 35–43. https://cran.r-project.org/web/packages/rworldmap/index.html

Stokman, F. N. (2001). Networks: Social. In *International Encyclopedia of the Social & Behavioral Sciences* (pp. 10509–10514). Elsevier. https://doi.org/10.1016/b0-08-043076-7/01934-3

Stone, G. W., & Lodhia, S. (2019). Readability of integrated reports: an exploratory global study. *Accounting, Auditing and Accountability Journal*, *32*(5), 1532–1557. https://doi.org/10.1108/AAAJ-10-2015-2275

Stricker, J., Chasiotis, A., Kerwer, M., & Günther, A. (2020). Scientific abstracts and plain language summaries in psychology: A comparison based on readability indices. *PLOS ONE, 15*(4), e0231160. https://doi.org/10.1371/journal.pone.0231160

The igraph core team. (2019). *igraph – Network analysis software* (1.2.4.1). https://igraph.org/

Traag, V., Waltman, L., & Eck, N. J. (2019). From Louvain to Leiden: guaranteeing well-connected communities. *Scientific Reports, 9*.

Uddin, S., & Khan, A. (2016). The impact of author-selected keywords on citation counts. *Journal of Informetrics, 10*(4), 1166–1177. https://doi.org/10.1016/j.joi.2016.10.004

Wasserman, S., & Faust, K. (2003). *Social Network Analysis: Methods and Applications*. Cambridge: University Press. https://books.google.com.co/books?hl=es&lr=&id=CAm2DpIqRUIC&oi=fnd&pg=PR21&ots=HxGowg0ARd&sig=VWM0wOXFrV5tlnbr9DdgJiBZFyc&redir_esc=y#v=onepage&q&f=false

Williams, L. S. (2008). The mission statement: A corporate reporting tool with a past, present, and future. *Journal of Business Communication, 45*(2), 94–119. https://doi.org/10.1177/0021943607313989

Yang, K., Fujisaki, I., & Ueda, K. (2020). Interplay of network structure and neighbour performance in user innovation. *Palgrave Communications, 6*(1), 1–8. https://doi.org/10.1057/s41599-019-0383-x

Yitzhaki, M. (1994). Relation of title length of journal articles to number of authors. *Scientometrics, 30*(1), 321–332. https://doi.org/10.1007/BF02017231

Yitzhaki, M. (1997). Variation in informativity of titles of research papers in selected humanities journals: A comparative study. *Scientometrics, 38*(2), 219–229. https://doi.org/10.1007/BF02457410

Yitzhaki, M. (2002). Relation of the title length of a journal article to the length of the article. *Scientometrics, 54*(3), 435–447. https://doi.org/10.1023/A:1016038617639

Yule, G. (1944). *The Statistical Study of Literary Vocabulary*. Cambridge University Press.

Zhang, H., Garrett, T., & Liang, X. (2015). The effects of innovation-oriented mission statements on innovation performance and non-financial business performance. *Asian Journal of Technology Innovation, 23*(2), 157–171. https://doi.org/10.1080/19761597.2015.1074495

**Tables and figures**

**Tables**

**Table 1 Journals' mission statement (JMS) and organisations' mission statement (MS)**

| Positive quality indicators/characteristics | Mission statement related-concept | Definition |
|---|---|---|
| Intuitive website | Purpose | "Journal scope of content is well-defined and aligns with journal title, article content and expectation of typical professional readers of the journal, instead of containing a broad scope of content." |
| Clear publisher information | Purpose; standards and behaviours | "The publisher is clearly identified and meets expected criteria for quality journals and publishing organizations." |
| Journal characteristics | Purpose; standards and behaviours; values; and strategy | "Journal affiliation and/or sponsorship is clearly identified and recognized as being an established scholarly society or academic institution. Content of journal articles aligns with journal content scope and meets the standards of the discipline." |

Source: the author based on Masten and Ashcraft (2017, p. 5) and Campbell (1989).

**Table 2 Summary of variables selected and groups**

| n=1,502 | | | | | | | |
|---|---|---|---|---|---|---|---|
| Type of variable | Variable | Percentage | | | | | |
| Factor | | | | | | | |
| | JMS | Overview | 12.7% | | | | |
| | | Aim, scope, other | 47.0% | | | | |
| | | Both | 40.3% | | | | |
| | Type of access | Open access status (DOAJ/ ROAD) | 13.3% | | | | |
| | | Non-registered | 86.7% | | | | |
| | Best quartile | Q1 | 37.5% | | | | |
| | | Q2 | 27.4% | | | | |
| | | Q3 | 22.8% | | | | |
| | | Q4 | 12.3% | | | | |
| Numeric | | Summary | | | | | |
| | | Min. | 1st Q | Median | Mean | 3rd Q | Max |
| | SJR | 0.1 | 0.22 | 0.42 | 0.81 | 0.88 | 17.97 |
| | H index | 1 | 11 | 20.5 | 34.63 | 47.75 | 283 |
| | Coverage (years) | 1 | 8 | 13 | 17.35 | 24 | 120 |

Source: the author based on SCImago journal ranking (2020), Scopus source list (2021) and journals website.

**Table 3 Summary statistics for FKGL and Yule's K**

| Index | Min. | 1st Q. | Median | Mean | 3rd. Q | Max |
|---|---|---|---|---|---|---|
| FKGL | 9.81 | 17.14 | 19.54 | 20.0 | 22.45 | 65.87 |
| Yule's K | 105.5 | 183.1 | 222.5 | 239.3 | 274.3 | 1250 |

Source: the author based on SCImago journal ranking (2020) and Scopus source list (2021). Processed with quanteda (Benoit et. al., 2020).

## Figures

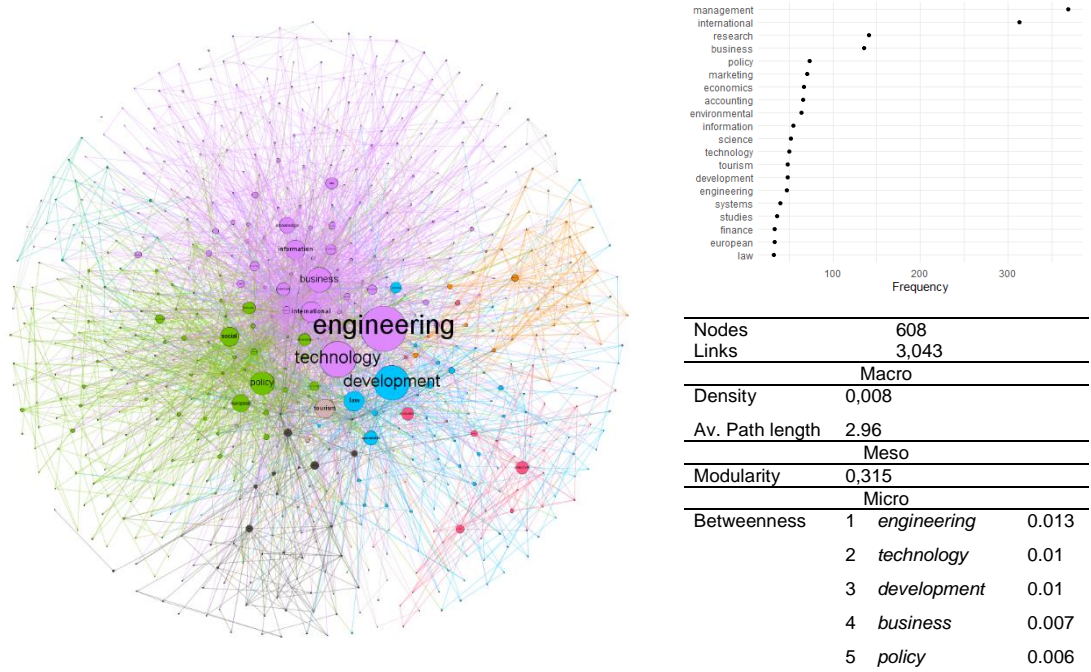

Figure 1 Semantic network – Titles (left side); top-20 most frequent tokens (top-right); macro, meso, micro (top-5) metrics table (bottom-right). Network layout algorithm: Fruchterman-Reingold (1991). Note: node and label size proportional to its betweenness score. Source: the author based on SCImago journal ranking (2020), Scopus source list (2021) and journals website. Processed with quanteda, igraph and gephi (Bastian et al., 2009; Benoit et. al., 2020; R Core Team, 2014; The igraph core team, 2019).

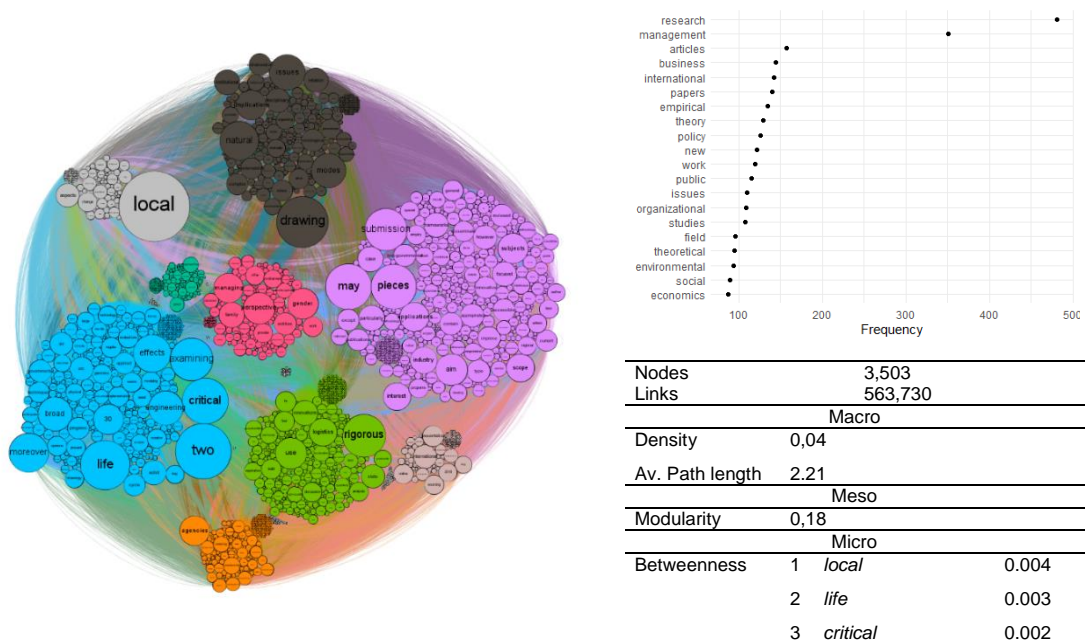

| | | |
|---|---|---|
| 4 | *rigorous* | 0.002 |
| 5 | *examining* | 0.002 |

**Figure 2 Semantic network – JMS top 10% in the SJR (left side); top-20 most frequent tokens (top-right); macro, meso, micro (top-5) metrics table (bottom-right). Network layout algorithm: Circular graph (Six & Tollis, 2006). Note: node and label size proportional to its betweenness score. Source: the author based on SCImago journal ranking (2020), Scopus source list (2021) and journals website. Processed with quanteda, igraph, and gephi (Bastian et al., 2009; Benoit et. al., 2020; R Core Team, 2014; The igraph core team, 2019).**

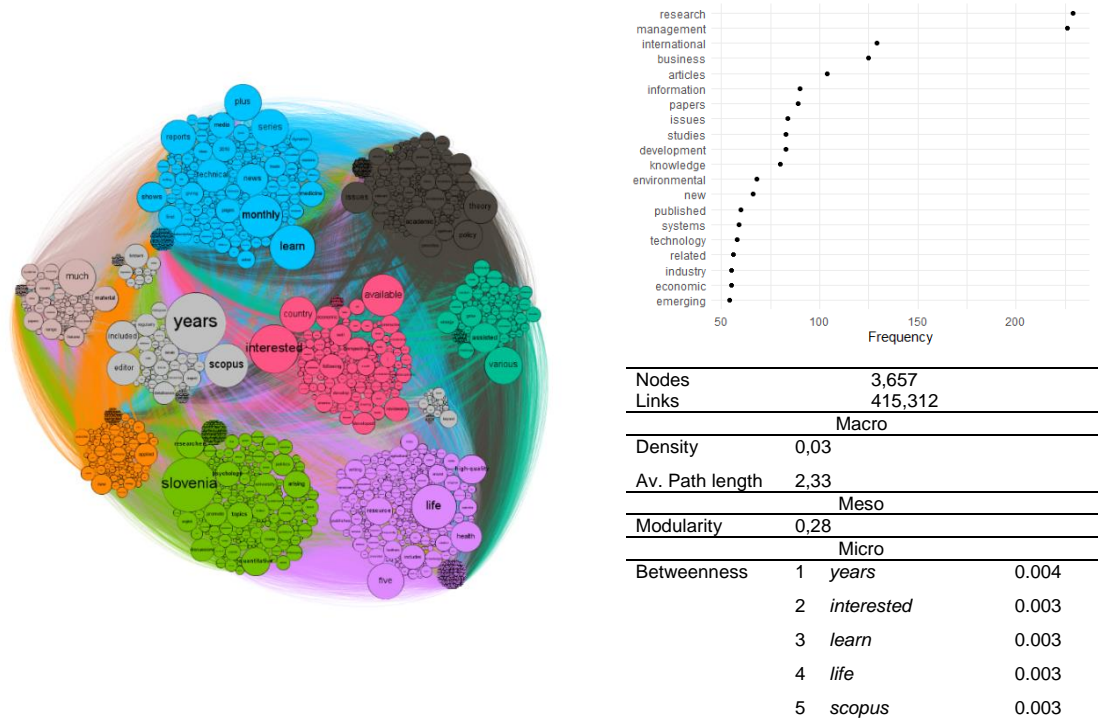

| | | | |
|---|---|---|---|
| Nodes | | 3,657 | |
| Links | | 415,312 | |
| | Macro | | |
| Density | 0,03 | | |
| Av. Path length | 2,33 | | |
| | Meso | | |
| Modularity | 0,28 | | |
| | Micro | | |
| Betweenness | 1 | *years* | 0.004 |
| | 2 | *interested* | 0.003 |
| | 3 | *learn* | 0.003 |
| | 4 | *life* | 0.003 |
| | 5 | *scopus* | 0.003 |

**Figure 3 Semantic network – JMS bottom 10% in the SJR (left side); top-20 most frequent tokens (top-right); macro, meso, micro (top-5) metrics table (bottom-right). Network layout algorithm: Circular graph (Six & Tollis, 2006). Note: node and label size proportional to its betweenness score. Source: the author based on SCImago journal ranking (2020), Scopus source list (2021) and journals website. Processed with quanteda, igraph and gephi (Bastian et al., 2009; Benoit et. al., 2020; R Core Team, 2014; The igraph core team, 2019).**

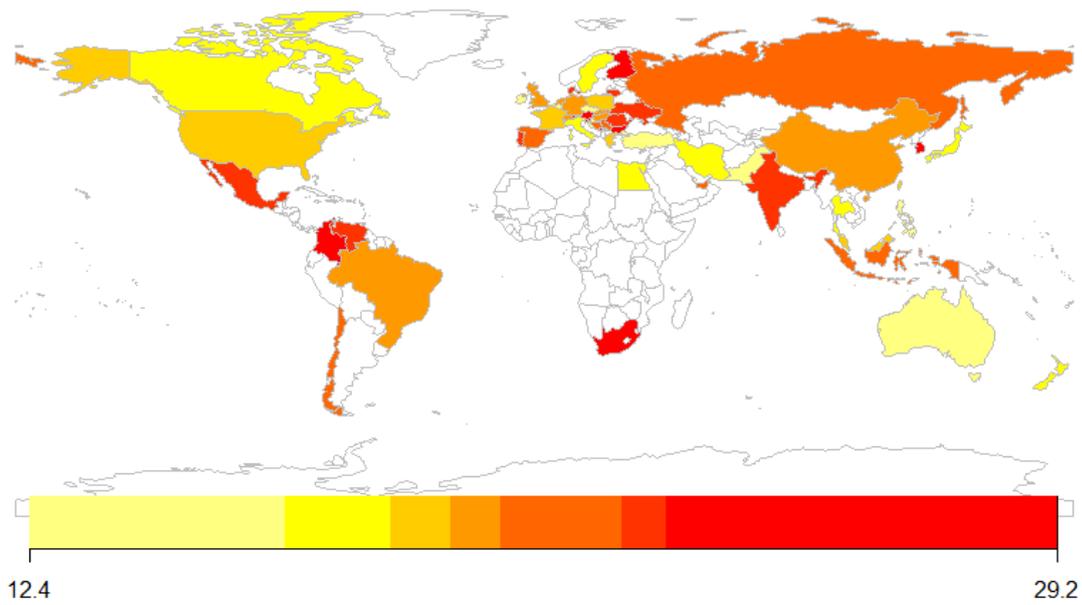

**Figure 4 World map – Median JMS FKGL according to publishers' countries. Source: the author based on SCImago journal ranking (2020) and Scopus source list (2021). Processed with quanteda, igraph, rworldmap, and gephi (Bastian et al., 2009; Benoit et. al., 2020; R Core Team, 2014; South, 2011; The igraph core team, 2019).**

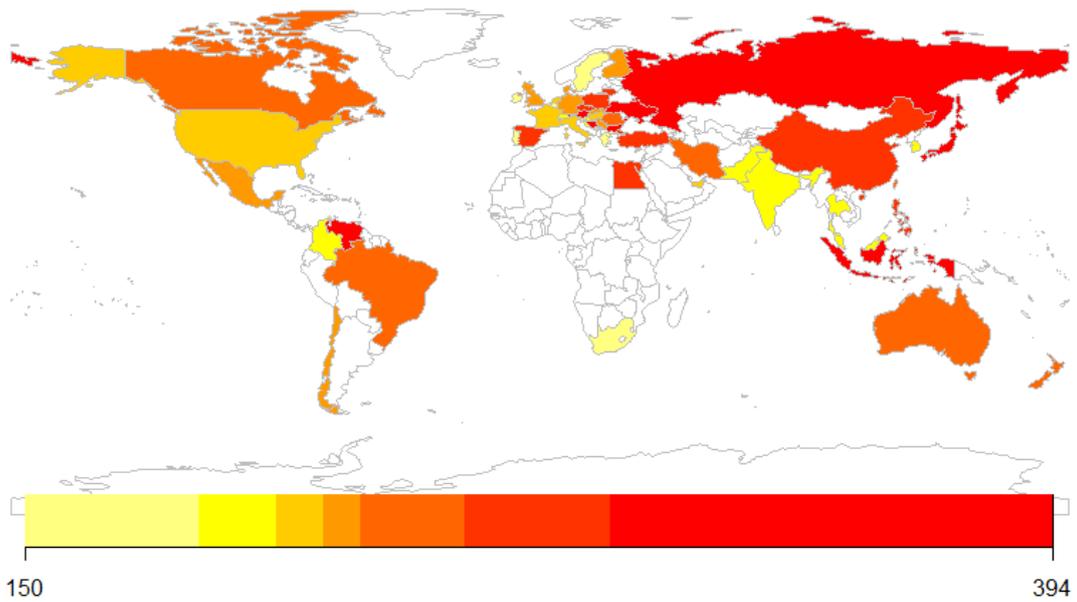

**Figure 5 World map – Median JMS Yule's K according to publishers' countries. Source: the author based on SCImago journal ranking (2020) and Scopus source list (2021). Processed with quanteda, igraph, rworldmap, and gephi (Bastian et al., 2009; Benoit et. al., 2020; R Core Team, 2014; South, 2011; The igraph core team, 2019).**

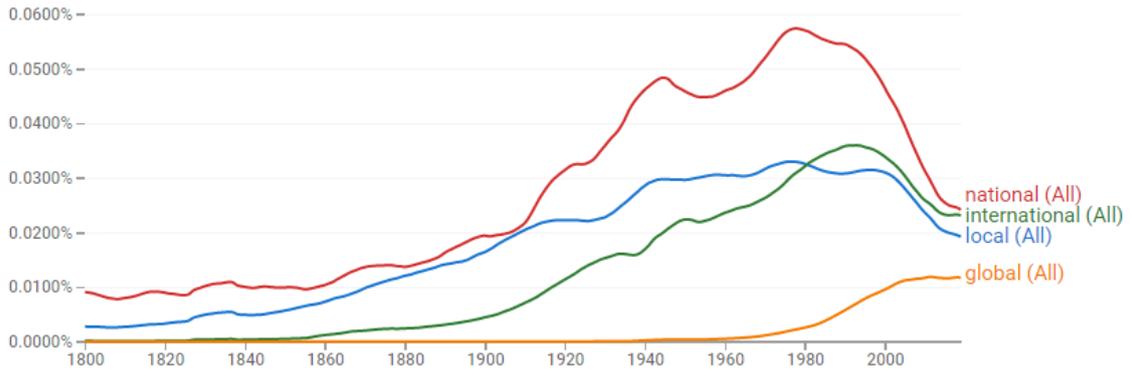

**Figure 6** Google's Books Ngram Viewer results for international and local. Source: Google (2021).